*submit stencil*

# Modeling the radio and optical/NIR afterglows of GRB 980703: a numerical study

KONG SiWei[1], HUANG YongFeng[1†], CHENG KwongSang[2], LU Tan[3]

[1] Department of Astronomy, Nanjing University, Nanjing 210093, China;
[2] Department of Physics, The University of Hong Kong, Pokfulam Road, Hong Kong, China
[3] Purple Mountain Observatory, Chinese Academy of Sciences, Nanjing 210008, China

**Extensive multi-band afterglow data are available for GRB 980703. Especially, its radio afterglow was very bright and was monitored until more than 1000 days after the trigger time. Additionally, there is no obvious special features, i.e., no re-brightenings, no plateau, and no special steep decay or slow decay in the multi-band afterglow light curves. All these conditions make GRB 980703 a precious sample in gamma-ray burst research. Here we use the observational data of GRB 980703 to test the standard fireball model in depth. It is found that the model can give a satisfactory explanation to the multi-band and overall afterglow light curves. The beaming angle of GRB 980703 is derived as ~ 0.23 radian, and the circum-burst medium density is ~ 27 cm$^{-3}$. The total isotropic equivalent kinetic energy of the ejecta is ~ 3.8 × 10$^{52}$ ergs. A rest-frame extinction of $A_V$ ~ 2.5 mag in the host galaxy is also derived.**

gamma ray bursts, jets and outflows

## 1   Introduction

Gamma-ray bursts (GRBs) are attractive astrophysical phenomena and had puzzled astronomers for a long time since their discovery in 1973[1]. The discovery of long-lived afterglow emission from GRBs, which spans from radio to X-ray and lasts from minutes to years, is a watershed in the field [2−4]. The detection of afterglow makes it possible for us to obtain broadband observational data, to identify the host galaxy, and to determine the redshift of GRBs. The so-called fireball model has been favored as the standard model and has been wildly discussed.

In the standard fireball model[5−11], the outflow of GRB, which moves relativistically, interacts with the surrounding medium to form an external shock. This shock will accelerate electrons to relativistic velocities. At the same time, a fraction of shock energy will be transferred to the magnetic field. These shock-accelerated relativistic electrons move in the magnetic field and emit synchrotron radiation. The resulting spectrum and light curve can be approximated as broken power-law functions[12], which can describe the main observed features of GRB afterglows well.

Received **?** **?**, 2009; accepted **?** **?**, 2009
doi:
†Corresponding author (email: hyf@nju.edu.cn)
Supported by the National Natural Science Foundation of China (Grant 10625313), by National Basic Research Program of China (973 Program 2009CB824800), and by a GRF of Hong Kong Government.

Within this framework, a lot of progress has been made on various aspects of GRB physics. For example, the total isotropic equivalent kinetic energies $E_0$ of many GRBs have been estimated; the geometry of the ejecta and the density structure of the surrounding medium have been revealed; and the values of some microphysics parameters of the shocks are studied in depth[13−15]. Also, valuable information about GRB host galaxies has been acquired by investigating the extinction law in the hosts[16, 17].

Today, although afteglows have been detected for several hundred GRBs, and although a lot of theoretical investigations have been done by many authors[13−15, 18, 19], detailed fitting to multi-band afterglows is still lacking, especially when the late afterglows are involved. To fit the overall afterglow light curves, we need to consider both the ultra-relativistic stage and the non-relativistic stage, and even the deep Newtonian stage[20], thus it can be done only numerically. To demonstrate the correctness of the fireball model, it is necessary to find a good GRB sample, which has multi-band and long-lasting afterglow data. We then can use this sample to test the fireball model in detail, by fitting the copious afterglow data.

In this paper, we present a detailed numerical study, by using the standard fireball model, on the multi-band afterglows of GRB 980703, which is special for having more than 1000 days of radio afterglow data at 1.43, 4.86 and 8.46 GHz and multi-band optical/NIR data from *K* to *B* bands. The outline of our paper is as follows. We review the observations of GRB 980703 in Section 2. In Section 3 we introduce our afterglow model, including the dynamics, the radiation process and other additional effects. Our numerical results and the fit to observations are presented in Section 4. Finally, Section 5 presents the conclusion and discussion.

## 2 Observations

GRB 980703 was detected on 1998 July 3.18 UT by the All-Sky Monitor (ASM) on the *Rossi X-ray Timing Explorer (RXTE)*[21], *Ulysses*[22], the Burst and Transient Source Experiment (BATSE) on the *Compton Gamma-Ray Observatory*[23] and the GRB monitor on *BeppoSAX*[24]. The BATSE data showed that the event is consisted of two pulses, each approximately lasting for 100 s, with a total duration of about 400 s, suggesting that it is a long-duration GRB[23]. The peak flux of the burst is 2.42 ± 0.06 photons cm$^{-2}$ s$^{-1}$ (50-300 keV; integrated over 1 s) and the fluence is (4.59 ± 0.42) × 10$^{-5}$ erg cm$^{-2}$ (> 20 keV) [23]. After that, comprehensive observations had been done by astronomers on detecting the afterglows of the burst.

About 22 hr after the burst, Galama et al.[25] observed the RXTE/ASM error box of GRB980703 by using the Narrow Field Instruments (NFI) on the *BeppoSAX* satellite and found a previously unknown X-ray source (1SAX J2359.1+0835), which was confirmed as the X-ray afterglow of GRB 980703 subsequently.

The discovery of radio and optical counterparts of GRB 980703 was first reported by Frail et al.[26] by using the VLA and Keck/Palomar respectively. The Keck observations suggested that the optical afterglow was at the position of $\alpha = 23^h 59^m 06^s.69$, $\delta = +08° 35' 07''.26$ (J2000.0), corresponding to the Galactic extinction of $A_V = 0.192$ mag and $E_{B-V} = 0.058$ mag[27]. More detailed observations and studies on GRB 980703 were subsequently obtained by many groups.

Bloom et al.[28] reported their optical/NIR observations of GRB 980703. They found a bright host galaxy with $R = 22.6$ mag. A rest-frame extinction of $A_V = 0.9 \pm 0.2$ mag was derived from the analysis of the broadband spectrum.

Castro-Tirado et al.[29] monitored the afterglow of GRB 980703 in both the R and H bands 22.5 hours after the burst. Their observations confirmed the power-law decay of the afterglow, after subtracting the contribution from the host galaxy with R = 22.49 ± 0.04 mag and H = 20.5 ± 0.25 mag. A reddened spectrum was also observed and a rest-frame extinction of $A_V$ = 2.2 mag was derived. This indicates that the burst may be surrounded by a dense medium.

Vreeswijk et al.[30] reported their X-ray and optical/NIR observations of GRB 980703. Their results are consistent with an extinction $A_V$ =1.50 ± 0.11 mag.

The host galaxy of GRB 980703 was also widely observed by astronomers at optical bands[31−34]. Several emission and absorption features were detected and a redshift of z = 0.966 was determined[31,32]. It is also found that the host galaxy is undergoing active star-formation. Three lines of the Balmer series were observed, which suggest the rest-frame extinction in the host galaxy as $A_V \approx$ 0.3 ± 0.3 mag[32].

Radio data of GRB 980703 were also obtained. Berger, Kulkarni & Frail[35] presented their radio observations from 350 to 1000 days after the burst. The host galaxy was discovered with flux densities of 39.3 ± 4.9, 42.1 ± 8.6 and 68.0 ± 6.6 μJy at 8.46, 4.86 and 1.43 GHz respectively. This is the first GRB host detection at radio wavelengths. They concluded that the burst locates in a region of star formation which is near the center of the host and the radio emission from the host is dominated by this region. The radio data from VLA observations before 250 days after the trigger were reported by Frail et al.[15].

In short, we believe that GRB 980703 is a good sample for us to study the standard fireball model. It has multi-band afterglow data, ranging from radio to optical wavelengths. The data are very complete in each band: in radio band, the observation spans from t ~1 day to t ~1000 day; and in optical bands, the observaion spans from t ≤ 1 day to t ~700 day. Additionally, there is no any special structures in the afterglow light curves, i.e, no rebrightenings, no plateau, no special steep decay or slow decay etc. It means that we do not need to consider many post-standard effects, such as the energy enjection, the density jump in the circum-burst medium, and so on. Below, we will use this event to check the correctness of the standard fireball model in depth.

## 3 Afterglow Model

### 3.1 Dynamics

We use the convenient equations developed by Huang et al.[20, 36−38] to describe the dynamics of the ejecta. The evolution of the bulk Lorentz factor ($\gamma$), the shock radius ($R$), the swept-up medium mass ($m$), and the half opening angle of the jet ($\theta$), are described by four differential equations,

$$\frac{d\gamma}{dm} = -\frac{\gamma^2 - 1}{m_{ej} + \varepsilon m + 2(1-\varepsilon)\gamma m} \tag{1}$$

$$\frac{dR}{dt} = \beta c \gamma \left(\gamma + \sqrt{\gamma^2 - 1}\right) \tag{2}$$

$$\frac{dm}{dR} = 2\pi R^2 (1-\cos\theta) n m_p \tag{3}$$

$$\frac{d\theta}{dt} = \frac{c_s(\gamma + \sqrt{\gamma^2-1})}{R} \tag{4}$$

with the comoving sound speed $c_s$ given by

$$c_s^2 = \hat{\gamma}(\hat{\gamma}-1)(\gamma-1)\frac{1}{1+\hat{\gamma}(\gamma-1)}c^2 \tag{5}$$

where $m_p$ is the mass of proton, $M_{ej}$ is the initial mass of the outflow, $n$ is the number density of the environment, $\beta = \sqrt{\gamma^2-1}/\gamma$, and $\hat{\gamma} \approx (4\gamma+1)/(3\gamma)$ is the adiabatic index[39]. $\varepsilon$ is the radiative efficiency, which equals 1 for the highly radiative case, and equals 0 in the adiabatic case. Equation (1) is a generic description of the deceleration of the external shock. It is applicable for both radiative and adiabatic shocks, and in both the ultrarelativistic and the nonrelativistic phases. Since the GRB ejecta becomes adiabatic several hours after the trigger, in this paper we assume the fireball is completely adiabatic all the time, so that $\varepsilon = 0$. In the simplest case, the interstellar medium (ISM) should be homogeneous. So, we take the number density ($n$) of the circum-burst environment of GRB 980703 as a constant value.

**3.2 Radiation Process**

The broadband afterglows are mainly from the synchrotron radiation of shocked electrons. But the synchrotron self-absorption is also unneglible, especially in radio bands[40]. As usual, we assume that the energy fraction of electrons and magnetic field with respect to the total energy are $\varepsilon_e$ and $\varepsilon_B$ respectively.

In the comoving frame, the synchrotron radiation power at frequency $\nu$ from a single electron with a Lorentz factor of $\gamma_e$ is given by[41]

$$P(\nu,\gamma_e) = \frac{\sqrt{3}e^3 B}{m_e c^2} F\left(\frac{\nu}{\nu_c}\right) \tag{6}$$

where $m_e$ is the mass of electron, $e$ is the electron charge, $B$ is the comoving magnetic field strength and $\nu_c = 3\gamma_e^2 eB/(4\pi m_e c)$. The function $F(x)$ is defined as

$$F(x) = x \int_x^{+\infty} K_{5/3}(k) dk \tag{7}$$

with $K_{5/3}(k)$ being the Bessel function. The total synchrotron radiation power from all the shock accelerated electrons, which may basically follow a power-law distribution $dN_e/d\gamma_e \propto (\gamma_e-1)^{-p}$ [20], at frequency $\nu$ is[41]

$$P_\nu = \int_{\min\{\gamma_{e,\min},\gamma_{e,c}\}}^{\gamma_{e,\max}} \frac{dN_e}{d\gamma_e} P(\nu,\gamma_e) d\gamma_e \tag{8}$$

Correspondingly, the self-absorption coefficient at frequency $\nu$ is[41]

$$k_\nu = \frac{p+2}{8\pi m_e \nu^2} \int_{\min\{\gamma_{e,\min},\gamma_{e,c}\}}^{\gamma_{e,\max}} \frac{dN_e}{d\gamma_e} \frac{1}{\gamma_e} P(\nu,\gamma_e) d\gamma_e \quad (9)$$

Self-absorption reduces the synchrotron radiation flux by a factor of $(1-e^{-\tau_\nu})/\tau_\nu$, where $\tau_\nu$ is the optical depth.

In some cases, inverse Compton scattering may also play a role in GRB afterglows[42, 43], but we ignore it because it usually does not affect the radio and optical/NIR afterglows significantly. In our calculations, we use the more realistic electron distribution function proposed by Huang & Cheng[20], which is correct even in the deep Newtonian phase.

### 3.3 Additional effects

Since the speed of light is finite, photons received by the observer at a particular time $t$ are not radiated simultaneously but come from a distorted ellipsoid[44], which is determined by

$$t = \int \frac{1-\beta\cos\Theta}{\beta c} dR \equiv \text{const} \quad (10)$$

within the jet boundaries, where $\Theta$ is the angle between the velocity of emitting material and the line of sight. This is the so-called equal arrival time surface (EATS). In our calculation, we integrate over the EATS to obtain an accurate observed flux.

Radio radiation from a compact extragalactic source will be scattered by the inhomogeneous interstellar plasma in our Galaxy. It will lead to large amplitude variations in the radio afterglow[45]. We use the method suggested by Walker[46, 47] to estimate the scintillation amplitude in our simulations. Scattering in the GRB host galaxy itself is not important as compared with the scattering in our Galaxy and can be negligible[45].

Dust extinction in GRB host galaxy and the Milk Way may reduce the observed optical/NIR flux. We use the extinction maps proposed by Schlegel, Finkbeiner & Davis[27] to correct for the extinction of our Galaxy. The extinction law in GRB host galaxy is quite uncertain and we will try to derive the extinction curve from our fit. The luminosity of GRB host galaxy can not be neglected and we also take it into account in our modeling.

## 4 Numerical Results

In this section we reproduce the radio and optical/NIR afterglows of GRB 980703, using the model described in Section 3. We take the radio data from Berger, Kulkarni & Frail[35] and Frail et al.[15], and the optical/NIR data from Bloom et al.[28, 48], Djorgovski et al.[31], Castro-Tirado et al.[29], Vreeswijk et al.[30], Holland et al.[33] and Sokolov et al.[34]. In our study, we use an assumptive cosmology of $H_0 = 65$ km s$^{-1}$ Mpc$^{-1}$, $\Omega_M = 0.30$ and $\Omega_\Lambda = 0.70$. Therefore, the redshift $z = 0.966$ corresponds to a luminosity distance of $D_L = 6.81 \times 10^6$ kpc (used to calculate flux density) and an angle distance of $D_A = 1.76 \times 10^6$ kpc (used to estimate scintillation amplitude). The parameter values of our best fit are presented in Table 1.

The observed radio afterglow light curves of GRB 980703 at 8.46, 4.86 and 1.43 GHz and our best fit are illustrated in Figure 1. The solid lines are our theoretical curves, and the dotted lines are the estimated scintillation amplitude. Contribution from the host is 39.3, 42.1 and 68.0 μJy at 8.46, 4.86 and 1.43 GHz respectively in our modeling, consistent with the observations of Berger et al.[35]. Generally speaking, the theoretical light curves predicted by the standard fireball model can explain the observed multi-band radio afterglows quite well during the whole observing span of $t \sim 1$ d — 1000 d. Note that the flux level at 1.43 GHz is much lower than that at 8.46 and 4.86 GHz. This is due to the more significant self-absorption at longer wavelength. Our calculations also show that the scattering effect should be more obvious at lower frequency due to the inhomogeneous interstellar plasma in our Galaxy. It is consistent with the observation, which reveals large amplitude variations at 1.43 GHz and 4.86 GHz.

Figure 2 illustrates the observed optical/NIR afterglow light curves of GRB 980703, and our best fit by using the same parameter values as in Figure 1. The observational data points have been corrected for the Galactic extinction[27]. The solid lines are our theoretical curves. Contribution from the host galaxy is included in our modeling. In most bands, the afterglow decreases to the level of the host galaxy after several tens of days. Interestingly, we see that the model gives a satisfactory fit to the light curve in all the seven optical/NIR bands.

GRB 980703 resides in its host galaxy. The extinction from the host galaxy is unnegligible. In plotting Figure 2, we have assumed different extinction magnitude in different bands, with the $V$-band extinction being $A_V = 2.5$ mag. We plot our extinction curve of GRB 980703 host in Figure 3. Comparing with the extinction laws of the local galaxies (i.e. the Large Magellanic Cloud, the Small Magellanic Cloud and the Milky Way)[49], we find that our extinction curve is most close to that of our Milky Way, as shown by the solid curve in Figure 3. However, note that our extinction coefficient at lower frequencies deviate from the solid curve significantly. This is not surprising. Similar phenomena have already been found by other researchers in recent studies[16, 17].

## 5 Discussion and Conclusion

GRB 980703 is a very important and attractive event. It has extensive broadband afterglow data. More importantly, its radio afterglow was very bright and monitored until more than 1000 days after the burst. Bright radio afterglows are rare in *Swift* era. This makes GRB 980703 one of the most valuable samples in GRB researches, especially in investigating the behavior of afterglows in the deep Newtonian phase[20]. Comparing with optical and X-ray observations, radio afterglows are distinctive in determining the circum-burst medium density[50] and the intrinsic kinetic energy[20].

In this paper, we chose GRB 980703 as a special sample to check the standard fireball model in depth. The overall dynamical evolution of the GRB ejecta is described by a generic dynamic equation, which has the virtue of being applicable for both radiative and adiabatic cases and in both ultrarelativistic and nonrelativistic phases. We assume that the GRB progenitor resides in the simplest environment, i.e., the ISM is homogeneous. We calculate the theoretical afterglow light curves numerically. It is found that the observed multi-band and long-lasting afterglow can be satisfactorily explained by the fireball model.

According to our calculations, the derived total isotropic equivalent kinetic energy in the afterglow phase is $3.8 \times 10^{52}$ ergs, and the half opening angle is 0.23 radian. It suggests that the intrinsic kinetic energy is $\sim 10^{51}$ ergs for a double-sided jet, consistent with the standard energy reservoir hypothesis[51]. The deduced ISM density of $n = 27.6$ cm$^{-3}$ is a typical value and is usual in the star formation regions. It is also consistent with the result of Djorgovski et al.[32], who have found that the host galaxy of GRB 980703 is undergoing active star formation. Note that the isotropic equivalent energy emitted in gamma rays is $E_\gamma \sim 6.01 \times 10^{52}$ ergs for GRB 980703[52]. The efficiency for producing gamma rays in the fireball, which is defined as $\eta_\gamma = E_\gamma/(E_\gamma+E_0)$, then can be derived as $\sim 60\%$. This value seems to be a bit too high. However, remember that we have omitted the highly radiative phase that lasts for a few hours since the GRB trigger. If we added the energy loss in the radiative phase into $E_0$, then $\eta_\gamma$ can be reduced significantly.

Our estimated extinction value is $A_V = 2.5$ mag. Although it is larger than some earlier estimation[28−30, 32−34], this value is still reasonable. In previous studies, the derived $A_V$ for GRB 980703 is highly dispersed, ranging from 0.3 to 2.2 . Currently it is difficult to determine which value is more exact. In the future, multi-band observations (especially including X-ray observations) will help to give a better constraint on the extinction of other GRB hosts. It is interesting to note that Rol et al.[53] have given a large lower limit of $A_V \approx 4.4$ mag for the host galaxy of GRB 051022.

Frail et al.[15] have analytically modeled the broadband afterglow of GRB 980703. However, their method is not ideal since they had to divide the afterglow light curve into a few segments, such as the ultra-relativistic phase and the non-relativistic phase. The light curve cannot transit smoothly from one segment to another segment in their calculations. In other words, their model cannot exactly describe the light curve behavior near the transition time. Our model is more realistic and is suitable for both the ultra-relativistic phase and the non-relativistic phase. It is correct even in the deep Newtonian phase[20]. Additionally, they did not include the EATS effect in their calculations. Thus our study is a more reliable test of the fireball model, especially at very late stages. However, note that in X-ray band, the afterglow light curve of GRB 980703 is poorly determined observationally, so that we can not give a meaningful comparison in X-rays. In the future, we will try to find better samples to carry out more detailed study.

1  Klebesadel R W, Strong I B, Olson R A. Observations of Gamma-ray bursts of cosmic origin. Astrophys J, 1973, 182: L85-L88
2  Costa E, Frontera F, Heise J, et al. Discovery of an X-ray afterglow associated with the $\gamma$-ray burst of 28 February 1997. Nature, 1997, 387: 783-785
3  van Paradijs J, Groot P J, Galama T, et al. Transient optical emission from the error box of the $\gamma$-ray burst of 28 February 1997. Nature, 1997, 386: 686-689
4  Frail D A, Kulkarni S R, Nicastro S R, et al. The radio afterglow from the $\gamma$-ray burst of 8 May 1997. Nature, 1997, 389: 261-263
5  Piran T. Gamma-ray bursts and the fireball model. Phys Rep, 1999, 314: 575-667
6  Mészáros P. Theories of gamma-ray bursts. Annl Rev Astron Astrophys, 2002, 40: 137-169
7  Piran T. The physics of gamma-ray bursts. Rev Mod Phys, 2004, 76: 1143-1210
8  Zhang B, Mészáros P. Gamma-ray bursts: progress, problems & prospects. Int J Mod Phys, A, 2004, 19: 2385-2472
9  Mészáros P. Gamma-ray bursts. Rep Prog Phys, 2006, 69: 2259-2322
10 Nakar, E. Short-hard gamma-ray bursts. Phys Rep, 2007, 442: 166-236
11 Zhang B. Gamma-ray burst in the Swift era. Chin J Astron Astrophys, 2007, 7: 1-50
12 Sari R, Piran T, Narayan R. Spectra and light curves of gamma-ray burst afterglows. Astrophys J, 1998, 497: L17-L20
13 Harrison F A, Yost S A, Sari R, et al. Broadband observations of the afterglow of GRB 000926: observing the effect of inverse Compton scattering. Astrophys J, 2001, 559: 123-130
14 Yost S A, Frail D A, Harrison F A, et al. The broadband afterglow of GRB 980329. Astrophys J, 2002, 577: 155-163
15 Frail D A, Yost S A, Berger E, et al. The broadband afterglow of GRB 980703. Astrophys J, 2003, 590: 992-998
16 Chen S L, Li A, Wei D M, Dust extinction of gamma-ray burst host galaxies: identification of two classes? Astrophys J, 2006, 647: L13-L16 17  Li Y, Li A, Wei D M, Determining the dust extinction of gamma-ray burst host galaxies: a direct method based on optical and X-ray photometry. Astrophys J, 2008, 678: 1136-1141
18  Panaitescu A, Kumar P. Jet energy and other parameters for the afterglows of GRB 980703, GRB 990123, GRB 990510, and GRB 991216 determined from modeling of multifrequency data. Astrophys J, 2001, 554: 667-677
19  Panaitescu A, Kumar P, Fundamental physical parameters of collimated gamma-ray burst afterglows. Astrophys J, 2001, 560: L49-L53
20  Huang Y F, Cheng K S. Gamma-ray bursts: optical afterglows in the deep Newtonian phase. Mon Not Roy Astron Soc, 2003, 341: 263-269
21  Levine A, Morgan E, Muto M. IAU Circ., 1998, 6966
22  Hurley K, Kouveliotou C. GRB 980703. GCN Circ., 1998, 125 (http://gcn.gsfc.nasa.gov/gcn/gcn3/125.gcn3)
23  Kippen R M. GRB 980703: BATSE observations. GCN Circ., 1998, 143 (http://gcn.gsfc.nasa.gov/gcn/gcn3/143.gcn3)
24  Amati L, Frontera F, Costa E, Feroci M. GRB 980703 – BeppoSAX/GRBM detection. GCN Circ., 1998, 146 (http://gcn.gsfc.nasa.gov/gcn/gcn3/146.gcn3)
25  Galama T J, van Paradijs J, Antonelli L A, et al. BeppoSAX NFI observations of GRB 980703. GCN Circ., 1998, 127 (http://gcn.gsfc.nasa.gov/gcn/gcn3/127.gcn3)
26  Frail D A, Halpern J P, Bloom J S, et al. GRB 980703: radio source/optical transient. GCN Circ., 1998, 128 (http://gcn.gsfc.nasa.gov/gcn/gcn3/128.gcn3)
27  Schlegel D J, Finkbeiner D P, Davis M. Maps of dust infrared emission for use in estimation of reddening and cosmic microwave background radiation foregrounds. Astrophys J, 1998, 500: 525-553
28  Bloom J S, Frail D A, Kulkarni S R, et al. The discovery and broadband follow-up of the transient afterglow of GRB 980703. Astrophys J, 1998, 508: L21-L24
29  Castro-Tirado A J, Zapatero-Osorio M R, Gorosabel J, et al. The optical/IR counterpart of the 1998 July 3 gamma-ray burst and its evolution. Astrophys J, 1999, 511: L85-L88
30  Vreeswijk P M, Galama T J, Owens A, et al. The X-ray, optical, and infrared counterpart to GRB 980703. Astrophys J, 1999, 523: 171-176
31  Djorgovski S G, Kulkarni S R, Goodrich R, et al. GRB 980703: Spectrum of the proposed optical counterpart. GCN Circ., 1998, 139 (http://gcn.gsfc.nasa.gov/gcn/gcn3/139.gcn3)
32  Djorgovski S G, Kulkarni S R, Bloom J S, et al. Spectroscopy of the host galaxy of the gamma-ray burst 980703. Astrophys J, 1998, 508: L17-L20
33  Holland S, Fynbo J P U, Hjorth J, et al. The host galaxy and optical light curve of the gamma-ray burst GRB 980703. Astron Astrophys, 2001, 371: 52-60
34  Sokolov V V, Fatkhullin T A, Castro-Tirado A J, et al. Host galaxies of gamma-ray bursts: spectral energy distributions and internal extinction. Astron Astrophys, 2001, 372: 438-455
35  Berger E, Kulkarni S R, Frail D A. The host galaxy of GRB 980703 at radio wavelengths – a nuclear starburst in an ultraluminous infrared galaxy. Astrophys J, 2001, 560: 652-658
36  Huang Y F, Dai Z G, Lu T. A generic dynamical model of gamma-ray burst remnant. Mon Not Roy Astron Soc, 1999, 309: 513-516
37  Huang Y F, Dai Z G, Lu T. On the optical light curves of afterglows from jetted gamma-ray burst ejecta: effects of parameters. Mon Not Roy Astron Soc, 2000, 316: 943-949
38  Huang Y F, Gou L J, Dai Z G, Lu T. Overall evolution of jetted gamma-ray burst ejecta. Astrophys J, 2000, 543: 90-96
39  Dai Z G, Huang Y F, Lu T. Gamma-ray burst afterglows from realistic fireballs. Astrophys J, 1999, 520: 634-640
40  Wu X F, Dai Z G, Huang Y F, Ma H T. Afterglow light curves of jetted gamma-ray burst ejecta in stellar winds. Chin J Astron Astrophys, 2004, 4: 455-472
41  Rybicki G B, Lightman A P. Radiative Processes in Astrophysics. 1979, New York: Wiley
42  Wei D M, Lu T. The influence of inverse Compton scattering on GRB afterglows: one possible way to flatten and steepen the light curves. Astron Astrophys, 2000, 360: L13-L16
43  Sari R, Esin A A. On the synchrotron self-Compton emission from relativistic shocks and its implications for gamma-ray burst afterglows. Astrophys J, 2001, 548: 787-799
44  Huang Y F, Lu Y, Wong A Y L, Cheng K S. A detailed study on the equal arrival time surface effect in gamma-ray burst afterglows. Chin J Astron Astrophys, 2007, 7: 397-404
45  Goodman J. Radio scintillation of gamma-ray burst afterglows. New Astron, 1997, 2: 449-460
46  Walker M A. Interstellar scintillation of compact extragalactic radio sources. Mon Not Roy Astron Soc, 1998, 294: 307-311

47  Walker M A. Erratum: Interstellar scintillation of compact extragalactic radio sources. Mon Not Roy Astron Soc, 2001, 321: 176
48  Bloom J S, Kulkarni S R. The hosts of GRB 980703 and GRB 971214. GCN Circ. 2000, 702 (http://gcn.gsfc.nasa.gov/gcn/gcn3/702.gcn3)
49  Pei Y C. Interstellar dust from the Milky Way to the Magellanic Clouds. Astrophys J, 1992, 395: 130-139
50  Frail D A, Cameron P B, Kasliwal M, et al. An energetic afterglow from a distant stellar explosion. Astrophys J, 2006, 646: L99-L102
51  Frail D A, Kulkarni S R, Sari R, et al. Beaming in gamma-ray bursts: evidence for a standard energy reservoir. Astrophys J, 2001, 562: L55-L58
52  Bloom J S, Frail D A, Sari R. The prompt energy release of gamma-ray bursts using a cosmological k-correction. Astron J, 2001, 121: 2879-2888
53  Rol E, van der Horst A, Wiersema K, et al. GRB 051022: physical parameters and extinction of a prototype dark burst. Astrophys J, 2007, 669: 1098-1106

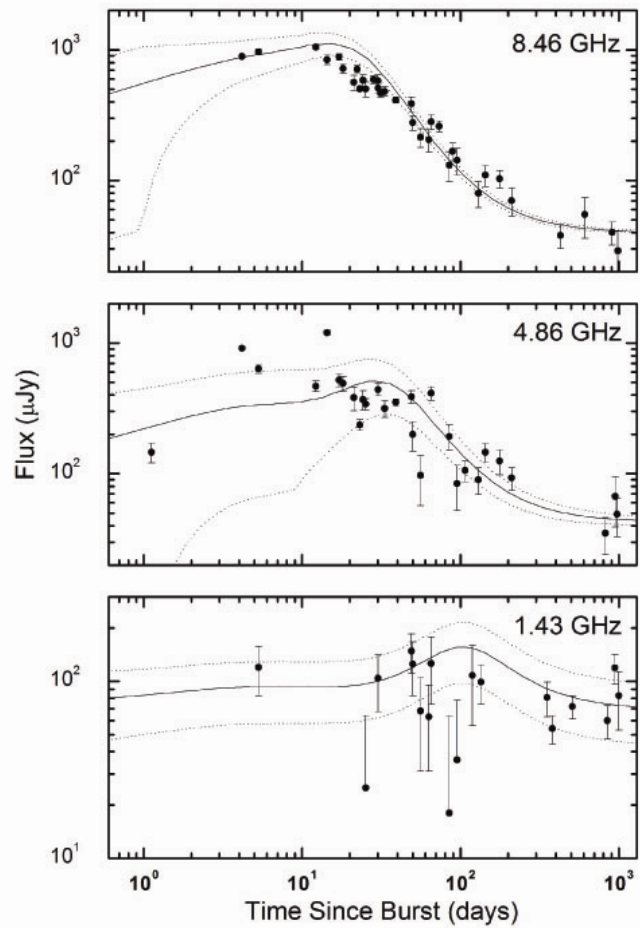

**Figure 1**  Our best fit to the multi-band radio afterglow light curves of GRB 980703. The observed data points are taken from Berger, Kulkarni & Frail[35] and Frail et al.[15]. The solid lines are our theoretical curves, and the dotted lines are estimated scintillation amplitudes. Contribution from the host has been added in our modeling.

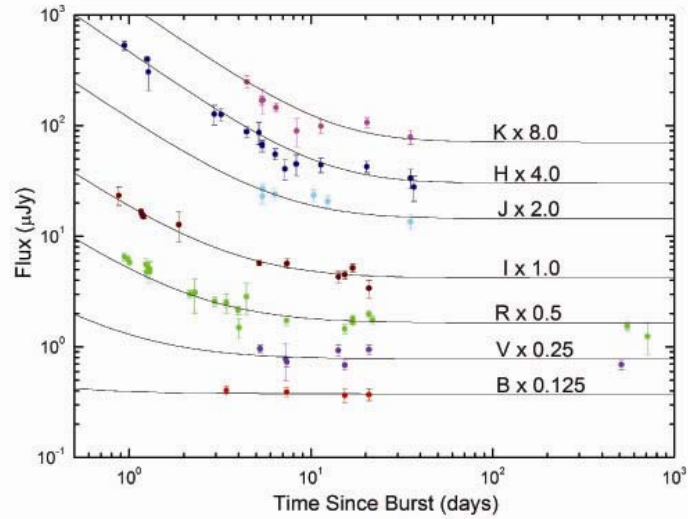

**Figure 2** Our best fit to the multi-band optical/NIR afterglow light curves of GRB 980703. The observed data points are taken from Bloom et al.[28, 48], Djorgovski et al.[31], Castro-Tirado et al.[29], Vreeswijk et al.[30], Holland et al.[33], and Sokolov et al.[34], and have been corrected for Galactic extinction[27]. The solid lines are our theoretical curves. Contribution from the host galaxy is included in our modeling.

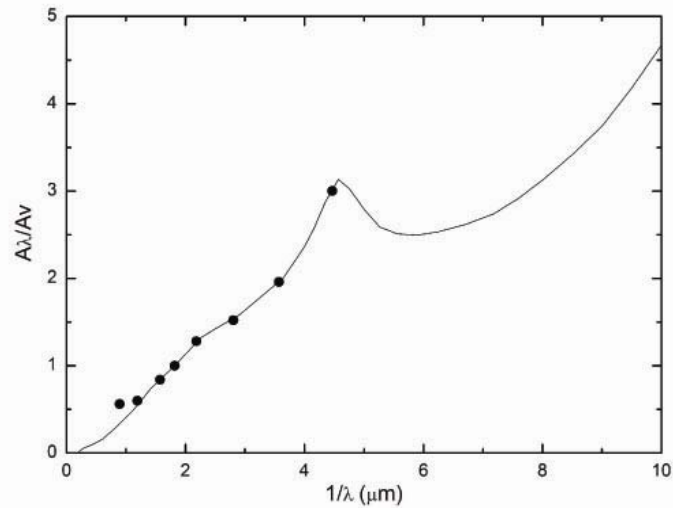

**Figure 3** The black dots illustrate our derived rest-frame extinction in the host galaxy of GRB 980703 along the line of sight (with $A_V$ = 2.5 mag). As a comparison, the solid line is the Galactic extinction curve proposed by Pei[49].

**Table 1** Modeling parameters of GRB 980703

| $E_0$ (ergs) | $\theta_0$ (radian) | $n$ (cm$^{-3}$) | $p$ | $\varepsilon_e$ | $\varepsilon_B$ | Host $A_V$ (mag) |
|---|---|---|---|---|---|---|
| $3.8 \times 10^{52}$ | 0.23 | 27.6 | 2.1 | 0.27 | $1.8 \times 10^{-3}$ | 2.5 |